\documentclass[twocolumn,showpacs,preprintnumbers,amssymb,pra]{revtex4}
\usepackage{graphicx}
\usepackage{rotating}
\usepackage{bm}
\usepackage{latexsym,epsfig}

\begin{document}

\title{Correlation Trends in the Hyperfine Structures of $^{210,212}$Fr} 
\author{$^1$B. K. Sahoo \footnote{Email: bijaya@prl.res.in}, $^1$D. K. Nandy, $^2$B. P. Das and $^3$Y. Sakemi}
\affiliation{Theoretical Physics Division, Physical Research Laboratory, Navrangpura, Ahmedabad-380009, India}
\affiliation{$^2$Theoretical Physics and Astrophysics Group, Indian Institute of Astrophysics, Bangalore 560034, India}
\affiliation{$^3$Cyclotron and Radioisotope Center, Tohoku University, Sendai, Miyagi 980-8578, Japan}
\date{Received date; Accepted date}

\begin{abstract}

We demonstrate the importance of electron correlation effects in the hyperfine structure 
constants of many low-lying states in $^{210}$Fr and $^{212}$Fr. This is achieved by calculating the magnetic dipole and electric 
quadrupole hyperfine structure constants using the Dirac-Fock approximation, second order many-body 
perturbation theory and the coupled-cluster method in the singles and doubles approximation in the relativistic 
framework. By combining our recommended theoretical results with the corresponding experimental 
values, improved nuclear magnetic dipole and electric quadrupole moments of the above isotopes are 
determined. In the present work, it is observed that there are large discrepancies between the hyperfine 
structure constants of the $7D_{5/2}$ state obtained from the experimental and theoretical studies whereas 
good agreements are found for the other $D_{5/2}$ states. Our estimated hyperfine constants for the $8P$, 
$6D$, $10S$ and $11S$ states could be very useful as benchmarks for the measurement of these quantities.
\end{abstract}

\maketitle

\section{Introduction}
 Francium (Fr) is the heaviest alkali atom in the periodic table possessing 87 electrons. Therefore, the properties of this system are expected to exhibit moderately strong correlation effects, and their determination calls for using powerful many-body 
methods. Like other alkali atoms, Fr atom is also 
being considered for many important experimental studies, prominent among them being the
measurement of the electric dipole moment (EDM) due to parity and time reversal symmetries \cite{sakemi, stancari} 
and parity nonconservation (PNC) effects due to neutral weak interaction \cite{stancari,dreischuh} and the nuclear 
anapole moment \cite{gomez1} owing to its relatively heavy size. Like the EDM and PNC interactions, the magnetic 
dipole hyperfine interaction has a fairly strong dependence on $Z$, the atomic number of the system 
\cite{sobelman} as it involves an electron interacting with the nucleus. Thus, theoretical investigations of hyperfine 
structures are necessary for EDM and PNC studies to test the accuracies of the wave functions in the 
nuclear region \cite{sahoo1,sahoo2}. On the other hand, comparison of theoretical results from various approximations 
with the experimental values can provide a comprehensive understanding of the passage of electron correlations from lower 
to higher levels of many-body theory. This knowledge is essential to validate the theoretical results when the 
experimental values are unavailable. Attempts have been made to investigate trends in the correlation effects in the 
calculations of hyperfine structure constants of the $S-$ states using lower order many-body methods \cite{das1,das2,
samii,owusu}, but such trends have not been demonstrated explicitly for states having higher angular momentum in Fr.

Here, we intend to show the variation in the trends of correlation effects in the evaluation of hyperfine structure 
constants of as many as 17 low-lying states in Fr considering relativistic second order many-body perturbation 
theory (MBPT(2) method) and the coupled-cluster (CC) method at various levels of approximation using the reference 
states obtained by the Dirac-Fock (DF) method with $V^{N-1}$ potential. In the present work, we have undertaken theoretical
studies of of the hyperfine structure constants of $^{210}$Fr and $^{212}$Fr isotopes for two reasons. First, $^{210}$Fr 
has been proposed as one of the most suitable Fr isotopes for both the PNC and EDM studies \cite{sakemi,stancari}. To 
draw meaningful conclusions and to be consistent in the findings from comparisons between the theoretical 
and experimental hyperfine structure constants of different states, it is necessary to consider the results for as many as states as 
possible for both the isotopes. The second, but the essential reason behind considering the above two isotopes is that experimental 
results of hyperfine structure constants for only a few selective low-lying states of $^{210}$Fr  
\cite{grossman,simsarian,grossman1,gomez} and some other excited states of $^{212}$Fr 
are available \cite{arnold}.

\section{Theory}
The Hamiltonian describing non-central form of hyperfine interaction between the electrons and nucleus in an
atomic system is expressed in terms of tensor operator products as \cite{schwartz}
\begin{eqnarray}
H_{hf} = \sum_k {\bf M}_n^{(k)} \cdot {\bf O}_{hf}^{(k)},
\end{eqnarray}
where ${\bf M}_n^{(k)}$ and ${\bf O}_{hf}^{(k)}$ are the spherical tensor operators with rank $k$ ($>0$) in the nuclear and
electronic coordinates respectively. Since these interaction strengths become much weaker with higher values of 
$k$, we consider only up to $k=2$ for the present case. Due to the coupling between the electronic (J) and 
nuclear (I) angular momentums, $|\gamma I J; F M_F \rangle$ states are the proper bases with total angular 
momentum ${\bf F}= {\bf J} + {\bf I}$ and corresponding azimuthal quantum number $M_F$ and $\gamma$ 
representing rest of the unspecified quantum numbers.

The energy splitting due to first order correction in the atomic state $|JM\rangle$ because of the hyperfine 
interaction is given by
\begin{eqnarray}
W_{F,J}^{(1)} &=& \langle \gamma IJ; FM_F|\sum_k {\bf M}_n^{(k)} . {\bf O}_{hf}^{(k)}|\gamma IJ; FM_F\rangle \nonumber \\
    &=& \sum_k (-1)^{I+J+F} \left  \{
                               \matrix
                                  {                                J & I & F \cr
                                I & J & k \cr
                                 } 
          \right \} \nonumber \\ && \times \langle I||M_n^{(k)}||I \rangle \langle J||O_{hf}^{(k)}||J \rangle 
\end{eqnarray}
which after expanding up to multipoles $k=2$, we get
\begin{eqnarray}
W_{F,J}^{(1)} &=& W_{F,J}^{M1} + W_{F,J}^{E2},
\end{eqnarray}
where $W_{F,J}^{M1}$ and $W_{F,J}^{E2}$ are the contributions due to the magnetic dipole (M1) with $k=1$ and 
electric quadrupole (E2) with $k=2$ interactions respectively. Traditionally it is expressed as
\begin{eqnarray}
W_{F,J}^{M1} &=& \frac{1}{2} A_{hf} K
\end{eqnarray}
and 
\begin{eqnarray}
W_{F,J}^{E2} &=& B_{hf} \frac{ \frac{3}{4} K^2 + \frac {3}{4} K- I(I+1)J(J+1)}{2I(2I-1)J(2J-1)} 
\end{eqnarray}
with the hyperfine structure constants defined as
\begin{eqnarray}
A_{hf}&=& \mu_N g_I \frac {\langle J||O_{hf}^{(1)}||J\rangle}{\sqrt{J(J+1)(2J+1)}} \label{eqna} 
\end{eqnarray}
and
\begin{eqnarray}
B_{hf}&=& 2 Q [ \frac {2J(2J-1)}{(2J+1)(2J+2)(2J+3)}]^{1/2} \nonumber \\ && \times \langle J||O_{hf}^{(2)}||J\rangle \label{eqnb},
\end{eqnarray}
where $\mu_N$ is the nuclear Bohr magneton, $g_I= \frac {\mu_I}{I}$, $\mu_I$ and $Q$ are the nuclear magnetic and quadrupole moments 
respectively,  and $K=F(F+1)-I(I+1)-J(J+1)$. It appears from the above expressions that the quantities 
$A_{hf}/g_I$ and $B_{hf}/Q$ are independent of nuclear factors, but these values can slightly vary among 
isotopes due to different nuclear potentials seen by electrons in these systems.

\section{Method of calculations}

In the Block's equation formalism, atomic wave function ($|\Psi_n \rangle$) of any of the considered states with 
closed-shell configuration $[6p^6]$ and a respective valence orbital $n$ in Fr can be expressed as \cite{lindgren}
\begin{eqnarray}
 |\Psi_n \rangle = \Omega_n |\Phi_n \rangle,
\end{eqnarray}
where $\Omega_n$ is known as the wave operator that generates virtual excitations from the reference state
$|\Phi_n \rangle$ to form all possible excited configurations to give rise the exact state $|\Psi_n \rangle$. 
The choice of $|\Phi_n \rangle$ can be crucial in obtaining accurate results, but we construct it using 
$V^{N-1}$ potential by expressing $|\Phi_n \rangle=a_n^{\dagger}|\Phi_0 \rangle$ with DF wave function 
$|\Phi_0 \rangle$ of the closed-core $[6p^6]$ obtained by considering the Dirac-Coulomb (DC) Hamiltonian ($H_{DC}$) 
and is the working reference state for all the considered atomic states. 

To account for the electron correlation effects through $\Omega_n$ in the evaluation of $|\Psi_n \rangle$, we 
proceed in two successive steps. In the first step, we consider correlation effects within the closed-shell 
configuration $[6p^6]$, following which we take into account correlation of the valence electron coupling with 
the core electrons to generate the virtual excitations. For this purpose, we express $\Omega_n$ as
\begin{eqnarray}
 \Omega_n = \Omega_0 + \Omega_n^v,
\end{eqnarray}
where $\Omega_0$ (independent of $n$) takes care of correlation effects from the closed-core while $\Omega_n^v$ 
is responsible for incorporating correlation corrections involving the electron from the valence orbital. In 
the perturbative expansion, we write \cite{lindgren}
\begin{eqnarray}
\Omega_n &=& \sum_{k=0}^2 \Omega_n^{(k)} 
\end{eqnarray}
with $\Omega_n^{(k)}$ corresponds to wave operator with presence of $k$ number of Coulomb interaction 
operator in it. The hyperfine structure constants in this method are determined by calculating the reduced
matrix elements of the $O_{hf}^{(1)}$ and $O_{hf}^{(2)}$ operators (below denoted simply by $O$) using the 
following expression
\begin{eqnarray}
\frac{\langle \Psi_n | O | \Psi_n \rangle} {\langle \Psi_n| \Psi_n \rangle}  
&=& \frac{\langle \Phi_n | \Omega_n^{\dagger} O \Omega_n | \Phi_n \rangle} {\langle \Phi_n| \Omega_n^{\dagger} 
\Omega_n | \Phi_n \rangle } \nonumber \\
&=& \frac{\langle \Phi_n | \sum_{i,j} \Omega_n^{(i) \dagger} O \Omega_n^{(j)} | \Phi_n \rangle} {\langle \Phi_n| \sum_{i,j} \Omega_n^{(i)\dagger} 
\Omega_n^{(j)} | \Phi_n \rangle }.
\end{eqnarray}
For the MBPT(2) method, $i+j \le 2$ in the above summations.

In the Fock space CC approach, we define $\Omega_0=e^T$ and $\Omega_n^v=e^T S_n $ that yields the form
\begin{eqnarray}
 |\Psi_n \rangle = e^T \{1+S_n \} |\Phi_n \rangle,
 \label{eqcc}
\end{eqnarray}
where $T$ and $S_n$ are the excitation operators involving core and core-valence electrons respectively.
Since our reference state is $|\Phi_0 \rangle$, we take $S_v$ in normal order with 
respect to $|\Phi_0 \rangle$ which, here, has been denoted by $\{ S_n\}$. It is worthwhile to mention that
$T$ is also in normal order with respect to $|\Phi_0 \rangle$ by construction. From a practical
point of view we only consider the singles and doubles excitations through the CC operators, known as the CCSD 
method in the literature \cite{szabo,bartlett}, by defining
\begin{eqnarray}
T = T_1 + T_2  \ \ \ \ \text{and} \ \ \ \ S_n = S_{1n} + S_{2n} .
\end{eqnarray}
Even though the CC operators have been approximated, we are able to account many important contributions from the
triples and quadruples through the non-linear terms of Eq. (\ref{eqcc}). In order to understand role of these
contributions in Fr for accurate evaluation of its hyperfine structure constants, we also determine
contributions only considering the linear terms of Eq. (\ref{eqcc}) by approximating
\begin{eqnarray}
 |\Psi_n \rangle = \{1+T_1+T_2+S_{1n} +S_{2n} \} |\Phi_n \rangle,
 \label{eqccl}
\end{eqnarray}
which we refer to as LCCSD method. One more remark on this approximation is that it resembles the form of the
singles and doubles configuration interaction (CISD) method. 

The hyperfine structure constants are determined in the CC method as
\begin{eqnarray}
\frac{\langle \Psi_n | O | \Psi_n \rangle} {\langle \Psi_n| \Psi_n \rangle}  
&=& \frac{\langle \Phi_n | \{1+ S_n^{\dagger}\} 
e^{T^{\dagger}} O e^T \{1+S_n\} | \Phi_n \rangle} {\langle \Phi_n| \{1+ S_n^{\dagger}\} 
e^{T^{\dagger}} e^T \{1+S_n\} | \Phi_n \rangle }. \nonumber \\
\label{prpeq}
\end{eqnarray}
The above expression has two non-truncative series $e^{T^{\dagger}} O e^T$ and $e^{T^{\dagger}} e^T$ in the 
numerator and denominator respectively. To account for the contributions that are significant from these series,
we have used the Wick's general theorem to divide these terms into effective one-body, two-body and three-body 
terms \cite{lindgren}. The effective one-body terms are dominant owing to the single particle form of the above 
$O$ operators, they are computed self-consistently and stored as intermediate parts before contracting with the 
corresponding $S_n$ operators. While the effective two-body and three-body terms are computed directly, but we 
also reuse the effective one-body terms to construct the effective two-body and three-body terms. To give 
an illustrate example of the adopted computational procedure in our calculations, we outline the important steps 
below. In the first step, we calculate the effective one-body terms of $e^{T^{\dagger}} e^T$ and store them as 
hole-hole (H-H), particle-particle (P-P), and particle-hole (P-H) blocks. The P-H block is first evaluated 
considering only linear terms and then stored as an intermediate effective P-H block which then contracted 
further with an extra $T_2^{\dagger}$ operator to get the H-P block. Consequently, the P-H block is obtained 
from the H-P block by multiplying appropriate phase factors and the procedure is repeated till the self-consistent results with tolerance size $10^{-8}$ 
achieved. Following this, the H-H and P-P blocks are constructed considering the direct contractions among the 
$T_2$ and $T_2^{\dagger}$ operators and along with the contractions of the H-P and H-P blocks with the $T_1$ 
operators. As a result, the H-H and P-P blocks still contain terms from infinite series through the H-P and P-H 
blocks. In a similar fashion we compute $e^{T^{\dagger}} O e^T$ but by slightly 
modifying the above strategy. Here we make use of the effective one-body terms of $e^{T^{\dagger}} e^T$
for the construction of the effective one-body diagrams of $e^{T^{\dagger}} O e^T$ and special attention has 
been paid to avoid the repetition of any diagram through the iterative procedure. We also replaced the 
$T_1$ and its conjugate operators appearing in the effective two-body and three-body terms of $e^{T^{\dagger}} O
e^T$ by the effective P-H block and H-P block of $e^{T^{\dagger}} e^T$ to improve the results with the 
contributions from higher order CC terms.

 We also estimate contributions from the important triple excitations considering them perturbatively in 
the above property evaluation expression by defining a triple excitation operator
\begin{eqnarray}
 S_{3n}^{pert} = \frac{1}{4} \sum_{ab,pqr} \frac{(H_{DC}T_2 + H_{DC} S_{2n})_{abn}^{pqr}}
 {\epsilon_n+ \epsilon_a+ \epsilon_b - \epsilon_p - \epsilon_q - \epsilon_r},  
\end{eqnarray}
where $a,b$ and $p,q,r$ indices represent for core and virtual orbitals respectively. Particles appearing in the 
subscripts are annihilated while those appear in the superscripts are created in the course of defining 
excitation processes. We refer to this approach as the CCSD$_{\text{t3}}$ method in this work.

\begin{table}
\caption{List of different parameters of bases used in the present calculations.}
 \begin{ruledtabular}
  \begin{tabular}{lccccc}
   &  $s$ & $p$ & $d$& $f$ & $g$  \\
  \hline
 & & & \\
 $N_l$  & 40 & 39 & 38 & 37 & 36 \\
 $\alpha_0$ & $2.0 \times 10^{-8}$ &  $2.5 \times 10^{-8}$ & $2.5 \times 10^{-8}$ & $2.1 \times 10^{-1}$ & $2.1 \times 10^{-7}$ \\
 $\beta$ & 5.06 & 5.04 & 5.06 & 5.08 & 5.15 \\
  \end{tabular}
 \end{ruledtabular}
 \label{tab1}
\end{table}
We use a recently developed basis function having the analytic exponential form with quadratic exponents to 
express the single particle wave functions. These functions for the orbitals having orbital angular 
momentum value $l$ are given as
\begin{eqnarray}
| \phi^l(r) \rangle &=& r^l \sum_{\nu =1}^{N_l} c_{\nu}^l e^{-\alpha_{\nu} r^4} |\chi(\theta, \varphi) \rangle,
\label{anbas}
\end{eqnarray}
where $|\chi(\theta, \varphi)\rangle $ represents the angular momentum part, $N_l$ corresponds to the total 
number of analytic functions considered in the calculations and $\alpha_{\nu}$ is an arbitrary coefficient 
constructed satisfying the even tempering condition between two parameters $\alpha_0$ and $\beta$ as 
\begin{eqnarray}
\alpha_{\nu} &=& \alpha_0 \beta^{\nu-1}.
\label{evtm}
\end{eqnarray}
The motive for using the above analytic function is to describe the atomic wave functions more accurately in the nuclear region
of a heavy atom like Fr than the Slater and Gaussian type orbitals. We give the list of 
parameters in Table \ref{tab1} used in the present calculations. 

\begin{table*}
\caption{Results for $A_{hf}/g_I$ and $B_{hf}/Q$ (both are in MHz) obtained using the DF, MBPT(2), LCCSD, CCSD and 
CCSD$_{\text{t3}}$ methods. Previously reported calculations using varieties of many-body methods are also quoted as ``Others''.} 
\begin{ruledtabular}
\begin{tabular}{lcccccc|cccccc}
  & \multicolumn{6}{c|}{$A_{hf}/g_I$ values}  & \multicolumn{6}{c}{$B_{hf}/Q$ values}  \\
  \cline{2-7} \cline{8-13} \\
State  & DF & MBPT(2) & LCCSD & CCSD & CCSD$_{\text{t3}}$ & Others & DF & MBPT(2) & LCCSD & CCSD & CCSD$_{\text{t3}}$ & Others \\
 \hline
 & & & \\
 $7s \ ^2S_{1/2}$  & 6531.06 & 10186.68&11457.29&9916.20  &9885.24   & 9927.69$^a$ \\
                   &         &         &        &         &          & 9947.07$^b$ \\
                   &         &         &        &         &          & 10155.40$^c$ \\
                   &         &         &        &         &          &  8432.65$^d$ \\
\vspace{0.4mm} 
 $7p \ ^2P_{1/2}$  & 707.70  & 1169.41  &1522.50 &1281.84  &1279.56   & 1308.67$^b$ \\
                   &         &         &        &         &          & 1265.77$^c$ \\
                   &         &         &        &         &          &  876.12$^e$ \\
\vspace{0.4mm} 
 $7p \ ^2P_{3/2}$  & 55.68   & 97.20   &124.80  &104.34   &104.28    & 103.38$^b$ & 118.03 & 234.51 &300.47  &258.99 &259.73 & 231$^e$ \\
                   &         &         &        &         &          & 115.09$^c$ \\
                   &         &         &        &         &          &  96.28$^e$ \\
 \vspace{0.4mm} 
 $6d \ ^2D_{3/2}$  & 36.24   & 44.10   &123.00  &85.02    &85.20     &            & 26.19  & 73.47  &100.17  &96.33  &98.61  \\
\vspace{0.4mm} 
 $6d \ ^2D_{5/2}$  & 14.28   & $-23.04$&$-84.36$  &$-58.92$ &$-58.82$ &            & 30.34  & 97.09  &134.01  &129.38 &131.40   \\
\vspace{0.4mm} 
 $8s \ ^2S_{1/2}$  & 1674.12 & 2234.40 &2337.92 &2159.16  &2151.90   & 2160.98$^a$ \\
                   &         &         &        &         &          & 2165.88$^b$ \\
                   &         &         &        &         &          & 2218.47$^c$ \\
                   &         &         &        &         &          & 2155.27$^d$ \\
\vspace{0.4mm} 
 $8p \ ^2P_{1/2}$  & 251.76  & 381.06  &457.38  &402.84   &402.01     &  408.68$^b$ \\
                   &         &         &        &         &          &  409.46$^c$ \\
\vspace{0.4mm} 
 $8p \ ^2P_{3/2}$  & 20.40   & 33.96   &40.62   &35.04    &35.03      &   34.25$^b$ &  43.29  &  78.74 &94.73   &83.81  &84.01 \\
                   &         &         &        &         &          &   39.64$^c$ \\
 \vspace{0.4mm} 
 $7d \ ^2D_{3/2}$  & 18.30   & 16.14   &36.45   &30.84    &30.90     &             &  13.23  &  24.99 &29.28   &29.97  &30.53 \\
\vspace{0.4mm} 
 $7d \ ^2D_{5/2}$  & 6.90    & $-12.24$&$-18.62$ &$-16.32$ &$-16.28$ &             & 14.57  &  32.84 &38.56   &39.45  &39.92  \\
\vspace{0.4mm} 
 $9s \ ^2S_{1/2}$  & 687.54  & 881.28  &910.01  &852.78   &849.72    &  852.39$^a$ \\
                   &         &         &        &         &          &  839.40$^d$ \\
\vspace{0.4mm} 
 $10s\ ^2S_{1/2}$  & 382.38  & 468.12  &449.63  &455.82   &454.26    & 402.97$^d$ \\
\vspace{0.4mm}
$8d \ ^2D_{3/2}$   &9.38    &7.75    &16.62   &14.45    &14.47     &            &6.79   &9.47   &13.15  &13.43  &13.66    \\
\vspace{0.4mm}
$8d \ ^2D_{5/2}$   &3.46     &$-5.34$    &$-7.62$   &$-6.90$    &$-6.88$     &            &7.35  &12.62  &17.26  &17.59  &17.78    \\
\vspace{0.4mm}
$9d \ ^2D_{3/2}$   &5.08     &4.22     &9.13    &7.84     &7.86      &            &3.68   &5.01   &7.08   &7.12   &7.24   \\
\vspace{0.4mm}
$9d \ ^2D_{5/2}$   &1.87     &$-3.02$    &$-3.89$   &$-3.53$    &$-3.52$     &            &3.96   &6.74   &9.21   &9.25   &9.35    \\
\vspace{0.4mm}
$11s \ ^2S_{1/2}$  &186.13   &238.72   &252.16  &231.13   &230.12    &  208.11$^d$ \\
\end{tabular}
\end{ruledtabular}
\label{tab2}
$^a$\cite{gomez}, $^b$\cite{safronova}, $^c$\cite{dzuba}, $^d$\cite{owusu}, $^e$\cite{martensson}.
\end{table*}

\begin{table*}[t]
\caption{Contributions from individual LCCSD and CCSD terms to the $A_{hf}/g_I$ calculations 
in the $7S$, $7P$, $6D$ and $8S$ states (in MHz). $Extra$ and $norm$ correspond to contributions from the non-linear 
terms of the CCSD method that are not mentioned explicitly and corrections due to the normalization of the wave 
functions respectively. $cc$ means complex conjugate.}
\begin{ruledtabular}
\begin{tabular}{lcccccccccccc}
CC term &  \multicolumn{2}{c}{$7s \ ^2S_{1/2}$} & \multicolumn{2}{c}{$7p \ ^2P_{1/2}$} &  
\multicolumn{2}{c}{$7p \ ^2P_{3/2}$} &\multicolumn{2}{c}{$6d \ ^2D_{3/2}$} 
& \multicolumn{2}{c}{$6d \ ^2D_{5/2}$}  & \multicolumn{2}{c}{$8s \ ^2S_{1/2}$} \\
  &  LCCSD  & CCSD &  LCCSD &  CCSD  & LCCSD & CCSD &  LCCSD & CCSD &  LCCSD &  CCSD  & LCCSD & CCSD \\
\hline        
& & \\
 $OT_1 + cc$                &$-440.401$ &$-437.580$&$-11.058$&$-18.540$&$-1.050$&$-1.740$&0.150    &0.042   &0.108    &0.066    &$-101.161$&$-101.52$ \\
\vspace{0.3mm}
$T_1^{\dagger} O T_1$       &7.422      &7.326     &0.042    &0.120    &0.006   &0.014   &$\sim 0$ &$\sim 0$&$\sim 0$ &$\sim 0$ &1.524     &1.536     \\
\vspace{0.3mm}
$T_1^{\dagger} O T_2$       &$-17.701$  &$-19.321$ &$-0.486$ &$-1.260$ &0.010   &0.018   &$\sim 0$ &$-0.002$&$-0.002$ &0.002    &$-4.092$  &$-4.230$  \\
\vspace{0.3mm}
$T_2^{\dagger} O T_2$       &$-72.300$  &$-75.721$ &$-4.548$ &$-5.628$ &1.848   &1.920   &1.230    &1.068   &2.202    &2.226    &$-16.356$ &$-16.548$ \\
\vspace{0.3mm}
$OS_{1v}+cc$                &3684.661   &2821.14   &486.901  &363.721  &31.956  &23.388  &57.438   &36.888  &18.901   &12.198   &494.341   &389.70    \\
\vspace{0.3mm}
$OS_{2v}+cc$                &1363.981   &1059.001  &225.720  &175.620  &28.698  &20.100  &$-5.064$ &$-4.518$&$-96.420$&$-70.021$&308.040   &246.00    \\
\vspace{0.3mm}
$T_1^{\dagger} OS_{2v}+cc$  &$-36.96$   &$-42.801$ &$-3.150$ &$-3.498$ &$-0.204$&$-0.228$&$-0.396$ &$-0.384$&$-0.144$ &$-0.138$ &$-4.320$  &$-10.098$ \\
\vspace{0.3mm}
$T_2^{\dagger}OS_{2v}+cc$   &$-227.890$ &$-191.930$&$-0.180$ &$-0.174$ &0.099   &$-0.060$&0.228    &0.126   &0.168    &0.036    &$-47.240$ &$-40.91$  \\
\vspace{0.3mm}
$S_{1v}^{\dagger}OS_{1v}$   &519.661    &304.620   &83.821   &46.758   &4.590   &2.460   &22.962   &9.462   &6.300    &2.544    &36.486    &22.680    \\
\vspace{0.3mm}
$S_{1v}^{\dagger}OS_{2v}+cc$&288.301    &178.501   &53.310   &32.268   &5.412   &2.976   &3.498    &1.014   &$-47.280$&$-23.706$&18.186    &13.518    \\
\vspace{0.3mm}
$S_{2v}^{\dagger}OS_{2v}$   &353.880    &225.481   &31.326   &21.690   &0.570   &0.618   &23.358   &13.482  &8.724    &4.476    &74.580    &49.398    \\
\vspace{0.3mm}
$Extra$                     &           &$-90.476$ &         &$-12.198$&        &0.522   &         &$-2.022$&         &$-2.208$ &          &$-56.382$ \\
\vspace{0.3mm}
$norm$                      &$-495.831$ &$-267.311$&$-46.836$&$-24.738$&$-2.604$&$-1.326$&$-16.596$&$-6.378$&8.778    &3.072    &$-91.911$ &$-46.723$ \\

\end{tabular} 
\end{ruledtabular}
\label{tab3}
\end{table*}

\section{Results and Discussion}

In order to demonstrate the propagation of electron correlation effects from the lower to higher orders of 
many-body methods, we present results using the DF, MBPT(2), LCCSD, CCSD and CCSD$_{t3}$ methods for 
$A_{hf}/g_I$ and $B_{hf}/Q$ of the first 17 low-lying states of $^{210}$Fr in Table \ref{tab2}. We find 
changes beyond third decimal places in these results for all the states except for the $A_{hf}/g_I$ 
values of the $s$-states, which are reduced by $-1.141$, $-0.242$, $-0.095$, $-0.047$ and $-0.028$ (in MHz) 
in the $7S$, $8S$, $9S$, $10S$ and $11S$ states, respectively, of $^{212}$Fr compared to the results of $^{210}$Fr.
We account for these differences later while considering the results for $^{212}$Fr. It can be seen from the 
above table that the MBPT(2) results are larger than the corresponding DF results for both the properties except 
for the $A_{hf}/g_I$ values of the $D_{5/2}$ states. In fact, there are sign differences between the DF and 
MBPT(2) results of $A_{hf}/g_I$ in the $D_{5/2}$ states implying that correlation effects are very strong for these 
states. It is found that the LCCSD method yields much larger values for both $A_{hf}/g_I$ and $B_{hf}/Q$ compared 
to the other methods. The non-linear terms in the CCSD methods contribute substantially but with opposite sign to 
reduce the LCCSD values. Triples through the CCSD$_{t3}$ method also further reduce the 
$A_{hf}/g_I$ values while they lead to small increases in the values of $B_{hf}/Q$. Thus, we conclude that the 
correlations effects represented by the non-linear terms that correspond to triple and quadruple 
excitations are crucial for the accurate evaluation of the hyperfine structure 
constants in Fr atom.
  
\begin{table*}[t]
\caption{Contributions from individual LCCSD and CCSD terms to the $A_{hf}/g_I$ calculations 
in the $8P$, $7D$, $9S$ and $10S$ states (in MHz). $Extra$ and $norm$ correspond to contributions from the non-linear 
terms of the CCSD method that are not mentioned explicitly and corrections due to the normalization of the wave 
functions respectively.}
\begin{ruledtabular}
\begin{tabular}{lcccccccccccc}
CC term &  \multicolumn{2}{c}{$8p \ ^2P_{1/2}$} &  \multicolumn{2}{c}{$8p \ ^2P_{3/2}$} & 
\multicolumn{2}{c}{$7d \ ^2D_{3/2}$} & \multicolumn{2}{c}{$7d \ ^2D_{5/2}$}  
& \multicolumn{2}{c}{$9s \ ^2S_{1/2}$}   & \multicolumn{2}{c}{$10s \ ^2S_{1/2}$} \\
&  LCCSD & CCSD &  LCCSD &  CCSD  & LCCSD & CCSD &  LCCSD &CCSD &  LCCSD &  CCSD  & LCCSD & CCSD \\
\hline        
& & \\
 $OT_1 + cc$                &$-3.090$ &$-6.198$ &$-0.264$ &$-0.564$ &0.084   &0.036    &0.054     &0.036     &$-40.098$&$-40.410$&$-19.895$&$-20.096$ \\ 
\vspace{0.3mm}
$T_1^{\dagger} O T_1$       &0.096    &0.036    &0.001    &0.001    &$\sim 0$&$\sim 0$ &$\sim 0$  &$\sim 0$  &0.582    &0.594    &0.285    &0.291  \\ 
\vspace{0.3mm}
$T_1^{\dagger} O T_2+cc$    &0.138    &$-0.276$ &0.001    &0.003    &$\sim 0$&$-0.001$ &$-0.001$  &$-0.001$  &$-1.626$ &-$1.680$ &$-0.807$ &$-0.833$ \\ 
\vspace{0.3mm}
$T_2^{\dagger} O T_2$       &$-1.764$ &$-0.252$ &1.164    &0.942    &0.720   &0.540    &1.032     &0.972     &$-7.476$ &$-7.524$ &$-3.846$ &$-3.872$ \\ 
\vspace{0.3mm}
$OS_{1v}+cc$                &121.861  &93.060   &8.670    &6.468    &11.700  &8.646    &3.606     &2.610     &162.780  &128.701  &74.151   &60.405 \\ 
\vspace{0.3mm}
$OS_{2v}+cc$                &71.940   &9.510    &9.666    &7.008    &1.524   &0.762    &$-31.716$ &$-25.458$ &122.341  &98.400   &60.806   &49.144 \\ 
\vspace{0.3mm}
$T_1^{\dagger} OS_{2v}+cc$  &$-1.074$ &$-1.212$ &$-0.072$ &$-0.084$ &$-0.150$&$-0.156$ &$-0.126$  &$-0.054$  &$-3.504$ &$-4.080$ &$-1.761$ &$-2.051$ \\ 
\vspace{0.3mm}
$T_2^{\dagger}OS_{2v}+cc$   &$-0.540$ &$-0.060$ &-0.036   &$-0.018$ &0.078   &0.048    &0.054     &0.030     &$-18.321$&$-15.962$&$-9.020$ &$-7.874$ \\ 
\vspace{0.3mm}
$S_{1v}^{\dagger}OS_{1v}$   &14.760   &8.616    &0.924    &0.510    &1.920   &1.050    &0.486     &0.252     &9.630    &6.018    &3.964    &2.630  \\ 
\vspace{0.3mm}
$S_{1v}^{\dagger}OS_{2v}+cc$&8.880    &5.814    &0.948    &0.588    &2.160   &1.218    &$-3.732$  &$-2.640$  &3.342    &2.940    &0.941    &1.124  \\ 
\vspace{0.3mm} 
$S_{2v}^{\dagger}OS_{2v}$   &9.540    &6.918    &0.306    &0.234    &6.084   &4.254    &2.526     &1.572     &29.508   &19.638   &14.666   &9.815  \\ 
\vspace{0.3mm}
$Extra$                     &         &$-6.408$ &         &0.030    &        &3.366    &          &$-1.584$  &         &23.568   &         &$-15.243$ \\ 
\vspace{0.3mm}
$norm$                      &$-15.276$&$-8.442$ &$-0.918$ &$-0.480$ &$-5.952$&$-3.024$ &2.250     &1.044     &$-34.747$&$-21.220$&$-16.646$&$-10.194$\\ 
\end{tabular} 
\end{ruledtabular}
\label{tab4}
\end{table*}

 We also mention about previously reported calculations of $A_{hf}/g_I$ and $B_{hf}/Q$ in Fr in Table 
 \ref{tab2} as ``Others'' \cite{owusu,gomez,safronova,dzuba,martensson}. Owusu et al. had investigated the 
core-polarization effects systematically in the $S-$ states of $^{212}$Fr using relativistic many-body perturbation
theory equivalent to our MBPT(2) method. These values are found to be smaller than our MBPT(2) results. SDpT method,
analogous to our LCCSD approximation but with important triples effects in the wave function 
determination, has been employed in Refs. \cite{gomez,safronova} to evaluate the $A_{hf}/g_I$ values of a few 
low-lying states and are in close agreement with our CCSD and CCSD$_{t3}$ results. Dzuba et al had employed 
a restricted Hartree-Fock method in the relativistic framework and incorporated correlation effects using
many-body perturbation theory to investigate correlation effects in the hyperfine structure constants
of a few low-lying states of $^{211}$Fr \cite{dzuba}. Heully and M{\aa}rtensson-Pendrill had employed a
relativistic many-body perturbation method treating polarization effects to all orders. Their results differ
\cite{martensson} significantly from our calculations.
 
\begin{table*}[t]
\caption{Contributions from individual LCCSD and CCSD terms to the $A_{hf}/g_I$ calculations 
in the $11S$, $8D$, and $9D$ states (in MHz). $Extra$ and $norm$ correspond to contributions from the non-linear 
terms of the CCSD method that are not mentioned explicitly and corrections due to the normalization of the wave 
functions respectively.}
\begin{ruledtabular}
\begin{tabular}{lcccccccccc}
CC term &  \multicolumn{2}{c}{$11s \ ^2S_{1/2}$} &  \multicolumn{2}{c}{$8d \ ^2D_{3/2}$} & 
\multicolumn{2}{c}{$8d \ ^2D_{5/2}$} & \multicolumn{2}{c}{$9d \ ^2D_{3/2}$}  
& \multicolumn{2}{c}{$9d \ ^2D_{5/2}$} \\
&  LCCSD & CCSD &  LCCSD &  CCSD  & LCCSD & CCSD &  LCCSD &CCSD &  LCCSD &  CCSD  \\
\hline        
& & \\
 $OT_1 + cc$                &$-10.583$  &$-10.704$   &0.048   &0.023    &0.030    &0.021     &0.024    &0.013    &0.016    &0.011    \\ 
\vspace{0.3mm}
$T_1^{\dagger} O T_1$       &0.150      &0.154       &$\sim 0$&$\sim 0$ &$\sim 0$ &$\sim 0$  &$\sim 0$ &$\sim 0$ &$\sim 0$ &$\sim 0$  \\ 
\vspace{0.3mm}
$T_1^{\dagger} O T_2+cc$    &$-0.429$   &$-0.443$    &$\sim 0$&$\sim 0$ &-0.001   &$\sim 0$  &$\sim 0$ &$\sim 0$ &$\sim 0$ &$\sim 0$ \\ 
\vspace{0.3mm}
$T_2^{\dagger} O T_2$       &$-2.111$   &$-2.113$    &0.642   &0.235    &0.518    &0.476     &0.240    &0.180    &0.277    &0.251    \\ 
\vspace{0.3mm}
$OS_{1v}+cc$                &44.080     &35.231      &4.044   &3.009    &1.259    &0.907     &2.082    &1.510    &0.650    &0.458    \\ 
\vspace{0.3mm}
$OS_{2v}+cc$                &32.902     &26.513      &1.290   &0.831    &$-14.290$&$-11.668$ &0.852    &0.570    &$-7.452$ &$-6.095$   \\ 
\vspace{0.3mm}
$T_1^{\dagger} OS_{2v}+cc$  &$-0.958$   &$-1.110$    &$-0.072$&$-0.076$ &$-0.025$ &$-0.027$  &$-0.036$ &$-0.041$ &$-0.013$ &$-0.014$   \\ 
\vspace{0.3mm}
$T_2^{\dagger}OS_{2v}+cc$   &$-4.100$   &$-4.225$    &0.006   &0.001    &0.014    &0.006     &0.004    &0.0002   &0.008    &0.003    \\ 
\vspace{0.3mm}
$S_{1v}^{\dagger}OS_{1v}$   &2.610      &1.667       &0.456   &0.253    &0.121    &0.063     &0.222    &0.118    &0.060    &0.030    \\ 
\vspace{0.3mm}
$S_{1v}^{\dagger}OS_{2v}+cc$&0.819      &0.776       &0.894   &0.553    &$-0.635$ &$-0.527$  &0.468    &0.292    &$-0.273$ &$-0.220$   \\ 
\vspace{0.3mm}  
$S_{2v}^{\dagger}OS_{2v}$   &8.063      &5.353       &2.580   &1.855    &1.124    &0.725     &1.350    &0.967    &0.598    &0.386    \\ 
\vspace{0.3mm}
$Extra$                     &           &$-6.099$    &        &$-1.608$ &         &$-0.341$  &         &$-0.850$ &         &$-0.210$   \\ 
\vspace{0.3mm}
$norm$                      &$-8.509$   &$-5.043$    &$-2.352$&$-1.273$ &0.795    &0.392     &$-1.174$ &$-0.602$ &0.366    &0.178    \\ 
\end{tabular} 
\end{ruledtabular}
\label{tab5}
\end{table*}

Having demonstrated that substantial correlation effects arise through the non-linear terms of the CCSD method above, 
it would be interesting to know whether these terms are more important for the accurate determination of atomic wave functions which 
are necessary for PNC and EDM studies in Fr atom or the non-linear terms appearing in Eq. (\ref{prpeq}). In case of the second possibility, it would be advisable to avoid large scale
computations in the evaluation of the amplitudes of the CC excitation operators. For this reason, we compare 
contributions arising term by term of Eq. (\ref{prpeq}) in the LCCSD and CCSD methods. These values 
for $A_{hf}/g_I$ are given in Tables \ref{tab3}, \ref{tab4} and \ref{tab5} for each considered state of 
$^{210}$Fr atom. Similarly, the calculated $B_{hf}/Q$ values are given in Tables \ref{tab6} and \ref{tab7}. 
Comparing contributions from individual terms of LCCSD and CCSD methods, we find the CC amplitudes are substantially 
changed by the correlation effects through the non-linear terms in the CCSD method. In fact, changes in the 
results due to normalization of the wave functions in both the methods (given as $norm$ in the above tables) 
are also quite large. Contributions given as $Extra$ are from the non-linear terms appearing in Eq. 
(\ref{prpeq}) and found to be comparatively small. Clearly, the consideration of non-linear terms of 
the CCSD method in the calculations of atomic wave functions are very crucial, and they should be considered for
obtaining high precision results and are very important for calculating quantities related to EDM and PNC studies 
in Fr. Even though Eq. (\ref{prpeq}) contains non-truncative series in the CCSD 
method, the linear terms contribute dominantly and are mainly responsible for determining the accuracies 
of the final results.

\begin{table*}[t]
\caption{Contributions from individual LCCSD and CCSD terms to the $B_{hf}/Q$ calculations (in MHz) in 
the $7P_{3/2}$, $8P_{3/2}$, $6D$ and $7D$ states. $Extra$ 
and $norm$ correspond to contributions from the non-linear terms of the CCSD method that are not mentioned 
explicitly and corrections due to the normalization of the wave 
functions respectively.}
\begin{ruledtabular}
\begin{tabular}{lcccccccccccc}
CC term &  \multicolumn{2}{c}{$7p \ ^2P_{3/2}$} & \multicolumn{2}{c}{$6d \ ^2D_{3/2}$} & 
\multicolumn{2}{c}{$6d \ ^2D_{5/2}$} &  \multicolumn{2}{c}{$8p \ ^2P_{3/2}$} 
& \multicolumn{2}{c}{$7d \ ^2D_{3/2}$} & \multicolumn{2}{c}{$7d \ ^2D_{5/2}$} \\
 &  LCCSD & CCSD & LCCSD &  CCSD  & LCCSD & CCSD &  LCCSD& CCSD & LCCSD & CCSD & LCCSD & CCSD \\

\hline        
& & \\
 $OT_1 + cc$                &$-2.237$ &$-3.690$ &0.112     &0.033    &0.234    &0.141     &$-0.564$ &$-1.202$ &0.064    &0.028    &0.123   &0.081  \\
\vspace{0.3mm}
$T_1^{\dagger} O T_1$       &0.010    &0.029    &$\sim 0$  &$\sim 0$ &$\sim 0$ &$\sim 0$  &0.002    &0.008    &$\sim 0$ &$\sim 0$ &$\sim 0$&$\sim 0$ \\
\vspace{0.3mm}
$T_1^{\dagger} O T_2$       &$-0.185$ &$-0.289$ &$-0.001$  &$-0.001$ &$\sim 0$ &$-0.001$  &$-0.051$ &$-0.094$ &$\sim 0$ &$\sim 0$ &$\sim 0$&$\sim 0$ \\
\vspace{0.3mm}
$T_2^{\dagger} O T_2$       &0.692    &0.888    &4.559     &4.332    &5.157    &4.926     &0.364    &0.248    &2.409    &2.060    &2.530   &2.162  \\
\vspace{0.3mm}
$OS_{1v}+cc$                &67.760   &49.601   &41.611    &26.721   &40.121   &25.521    &18.390   &13.72    &8.487    &6.271    &7.665   &5.542  \\
\vspace{0.3mm} 
$OS_{2v}+cc$                &103.331  &88.490   &26.401    &39.060   &55.031   &68.220    &31.631   &27.34    &5.851    &10.56    &14.671  &19.77  \\
\vspace{0.3mm}
$T_1^{\dagger} OS_{2v}+cc$  &$-0.434$ &$-0.482$ &$-0.289$  &$-0.277$ &$-0.306$ &$-0.300$  &$-0.156$ &$-0.176$ &$-0.109$ &$-0.115$ &$-0.112$&$-0.120$ \\
\vspace{0.3mm}
$T_2^{\dagger}OS_{2v}+cc$   &$-2.082$ &$-1.745$ &$-5.093$  &$-3.812$ &$-6.073$ &$-4.504$  &$-0.677$ &$-0.576$ &$-1.721$ &$-1.428$ &$-2.027$&$-1.652$ \\
\vspace{0.3mm}
$S_{1v}^{\dagger}OS_{1v}$   &9.737    &5.216    &16.649    &6.858    &13.385   &5.409     &1.957    &1.090    &1.394    &0.760    &1.041   &0.543  \\
\vspace{0.3mm}
$S_{1v}^{\dagger}OS_{2v}+cc$&14.081   &9.057    &0.240     &5.167    &8.190    &10.101    &2.534    &1.702    &2.830    &1.561    &3.293   &1.887  \\
\vspace{0.3mm}
$S_{2v}^{\dagger}OS_{2v}$   &$-0.578$ &$-0.436$ &3.294     &1.293    &1.873    &0.595     &0.153    &0.103    &1.620    &1.015    &1.472   &0.954  \\
\vspace{0.3mm}
$Extra$                     &         &$-1.786$ &          &$-0.006$ &         &$-0.045$  &         &0.010    &         &0.010    &        &0.003  \\
\vspace{0.3mm} 
$norm$                      &$-6.277$ &$-3.312$ &$-13.514$ &$-7.207$ &$-13.941$&$-7.415$  &$-2.138$ &$-1.162$ &$-4.781$ &$-2.961$ &$-4.666$&$-2.890$ \\

\end{tabular} 
\end{ruledtabular}
\label{tab6}
\end{table*}

 To understand the role of various correlation effects in the states belonging to different angular momenta, 
we discuss here the trends in the results through various CC terms for both the LCCSD and CCSD methods. It can
be seen from these tables that maximum contribution to $A_{hf}/g_I$ come from $OS_{1v}$ and its complex 
conjugate ($cc$) terms, which correspond to all order pair-correlation effects \cite{sahoo1,sahoo3}, followed by 
$OS_{2v}+cc$ terms, which represent the all order core-polarization effects \cite{sahoo1,sahoo3}, in the $S$, 
$P_{1/2}$ and $D_{3/2}$ states implying that the pair-correlation contributions are very important in all these 
cases. In the $P_{3/2}$ states, contributions from both these CC terms are almost same in magnitude while 
$OS_{2v}+cc$ are the dominant contributors in the $D_{5/2}$ states. We also find that the trends of 
contributions in the evaluation of $B_{hf}/Q$ are different from those of $A_{hf}/g_I$.  The core 
contributions coming through the excitation of the $T$ operators are comparatively significant for the $S-$ 
states. In general contributions through the single core excitations (generated by $T_1$) are found to be 
important. However, there are large core contributions through the $T_2$ operator, particularly, for 
the ground state. 

\begin{table*}[t]
\caption{Contributions from individual LCCSD and CCSD terms to the $B_{hf}/Q$ calculations (in MHz) 
in the $8D$ and $9D$ states. $Extra$ 
and $norm$ correspond to contributions from the non-linear terms of the CCSD method that are not mentioned 
explicitly and corrections due to the normalization of the wave 
functions respectively.}
\begin{ruledtabular}
\begin{tabular}{lcccccccc}
CC term & \multicolumn{2}{c}{$8d \ ^2D_{3/2}$} & \multicolumn{2}{c}{$8d \ ^2D_{5/2}$} &  \multicolumn{2}{c}{$9d \ ^2D_{3/2}$} 
& \multicolumn{2}{c}{$9d \ ^2D_{5/2}$} \\
 &  LCCSD & CCSD & LCCSD &  CCSD  & LCCSD & CCSD &  LCCSD& CCSD \\

\hline        
& & \\
 $OT_1 + cc$                &0.035     &0.017     &0.065     &0.044     &0.019     &0.010      &0.036    &0.024   \\
\vspace{0.3mm}
$T_1^{\dagger} O T_1$       & $\sim 0$ & $\sim0$  & 0.0001   &0.0001    & $\sim 0$ & $\sim0$   & 0.0001  & $\sim 0$ \\
\vspace{0.3mm}
$T_1^{\dagger} O T_2$       & $-0.0001$&$-0.0002$ & $-0.0001$& $-0.0002$& $-0.0001$& $-0.0001$ & $\sim 0$& $-0.0001$ \\
\vspace{0.3mm}
$T_2^{\dagger} O T_2$       &1.214     &0.992     &1.262     &1.035     &0.653     &0.523      &0.677    &0.545   \\
\vspace{0.3mm}
$OS_{1v}+cc$                &2.938     &2.185     &2.675     &1.926     &1.514     &1.096      &1.381    &0.972   \\
\vspace{0.3mm}  
$OS_{2v}+cc$                &2.429     &4.490     &6.338     &8.560     &1.226     &2.270      &3.246    &4.354   \\
\vspace{0.3mm}
$T_1^{\dagger} OS_{2v}+cc$  &$-0.051$  &$-0.055$  &$-0.053$  &$-0.057$  &$-0.028$  &$-0.029$   &$-0.028$ &$-0.030$  \\
\vspace{0.3mm}
$T_2^{\dagger}OS_{2v}+cc$   &$-0.811$  &$-0.682$  &$-0.953$  &$-0.786$  &$-0.429$  &$-0.360$   &$-0.503$ &$-0.415$  \\
\vspace{0.3mm}
$S_{1v}^{\dagger}OS_{1v}$   &0.351     &0.183     &0.257     &0.133     &0.162     &0.085      &0.127    &0.063   \\
\vspace{0.3mm}
$S_{1v}^{\dagger}OS_{2v}+cc$&1.350     &0.664     &1.364     &0.663     &0.756     &0.372      &0.748    &0.361   \\
\vspace{0.3mm}
$S_{2v}^{\dagger}OS_{2v}$   &0.781     &0.517     &0.757     &0.515     &0.433     &0.287      &0.425    &0.291   \\
\vspace{0.3mm}
$Extra$                     &          &$-1.667$  &          &$-1.791$  &          &$-0.815$   &         &$-0.871$  \\
\vspace{0.3mm}
$norm$                      &$-1.857$  &$-1.197$  &$-1.801$  &$-1.157$  &$-0.911$  &$-0.571$   &$-0.865$ &$-0.544$  \\

\end{tabular} 
\end{ruledtabular}
\label{tab7}
\end{table*}

We list the calculated and experimental values of $A_{hf}$ and $B_{hf}$ for $^{210}$Fr and $^{212}$Fr in 
Table \ref{tab8} for which a few high precision measurements are available. Coc et al. had 
obtained the $A_{hf}$ value for the ground $7S$ state as $7195.1(4)$ MHz using high-resolution spectroscopy 
\cite{coc}. In the same work, Coc et al had also given $A_{hf}$ and $B_{hf}$ values of the $7P_{3/2}$ state as 
$78.0(2)$ MHz and 51(4) MHz respectively. Using the two-photon excitation spectroscopy of $^{210}$Fr atom 
confined and cooled in a magneto-optical trap, Simsarian et al. had obtained the experimental $A_{hf}$ value 
for the $8S$ state as $1577.8(23)$ MHz at the ISOLDE facility \cite{simsarian}. Fairly recently, Gomez
et al. measured the $A_{hf}$ value for the $9S$ state as $622.25(36)$ MHz using a similar experimental 
technique \cite{gomez}. From another experiment, Coc. at al. have reported the $A_{hf}$ value of the $7P_{1/2}$
state as $945.6(5.8)$ MHz \cite{coc1}. A more precise value of $A_{hf}$ in the $7P_{1/2}$ state has been 
obtained by Grossman et al. as $946.6(5.8)$ MHz \cite{grossman}. Soon after that Grossman et al. extracted the 
$A_{hf}$ values for the $7D_{3/2}$ and $7D_{5/2}$ states by measuring hyperfine splittings as $22.3(5)$ MHz
and $-17.8(8)$ MHz respectively \cite{grossman1}. Here, they had neglected $B_{hf}$ value of the $7D_{3/2}$ 
state while estimating a large $B_{hf}$ value of 64(17) MHz for the $7D_{5/2}$ state. Similarly, for the $^{212}$Fr isotope 
there are many experimental results available for the $A_{hf}$ and $B_{hf}$ values measured using various 
spectroscopic techniques and are presented in the same Table \ref{tab8}. For instance Coc et al., adopting the
same experimental technique as for $^{210}$Fr, have measured the $A_{hf}$ values of the $7S$ and $7P_{1/2,3/2}$ 
states in $^{212}$Fr \cite{coc,coc1}. It can be noticed from the tabulated results for the $A_{hf}$ values 
that there is excellent agreement between the present calculations with the experimental values of the $7P_{1/2,3/2}$ 
states whereas our results differ from the measurements by about 1.2\% for the ground state. In another 
experiment Duong et al. had used stepwise laser excitation in collinear geometry with the on-line mass separator 
of the ISOLDE facility at CERN and measured the $A_{hf}$ and $B_{hf}$ values for the $7S$, $7P$ and $8P$ 
states of $^{212}$Fr \cite{duong}. These experimental results are also in good agreements with our 
calculations for the above states. Arnold et al. had further extended this project to carry out 
measurements of the hyperfine structure constants of the $10S$, $11S$, $8D$ and $9D$ states in $^{212}$Fr 
\cite{arnold}, which are also given in Table \ref{tab8}. It can be seen that these results and our calculations
agree very well.
 
\begin{table*}
\caption{Comparison between the theoretically determined and experimentally available $A_{hf}$ and $B_{hf}$ 
results (in MHz) of $^{210}$Fr and $^{212}$Fr.}
\begin{ruledtabular}
 \begin{tabular}{lcccc c cccc}
    &  \multicolumn{4}{c}{$^{210}$Fr}  &  & \multicolumn{4}{c}{$^{212}$Fr} \\
\cline{2-5} \cline{7-10} \\
\multicolumn{1}{c}{States}   & \multicolumn{2}{c}{$A_{hf}$} &  \multicolumn{2}{c}{$B_{hf}$} & 
& \multicolumn{2}{c}{$A_{hf}$} & \multicolumn{2}{c}{$B_{hf}$}\\
\cline{2-3} \cline{4-5} \cline{7-8} \cline{9-10}  \\
        & Present & Experiment &  Present & Experiment  &  &  Present & Experiment &  Present & Experiment \\
  \hline
& &  \\
$7s \ ^2S_{1/2}$  & 7253.45  &7195.1(4)$^a$  &   &      &   &  9123.72   & 9064.2(2)$^a$   &  &   \\
                  &          &               &   &      &   &            & 9064.4(1.5)$^b$ &  & \\
\vspace{0.4mm}
$7p \ ^2P_{1/2}$  & 938.90   &945.6(5.8)$^c$ &  &       &   &  1181.12   & 1189.1(4.6)$^c$ &  &   \\
\vspace{0.4mm}
                  &          &946.3(2)$^d$   &  &       &   &            & 1187.1(6.8)$^b$ &  &    \\
                  &          &               &  &       &   &            & 1192.0(2)$^d$   &  &    \\
\vspace{0.4mm}
$7p \ ^2P_{3/2}$  & 76.52    &78.0(2)$^a$    &  50.91   &51(4)$^a$ &   & 96.26   &97.2(1)$^a$ & $-25.97$ & $-26.0(2)^b$\\
                  &          &               &          &          &   &         &97.2(1)$^b$ &        &     \\     
\vspace{0.4mm}
$6d \ ^2D_{3/2}$  & 62.52    &               &  19.33   &          &   & 78.65  &  & $-9.86$  & \\
\vspace{0.4mm}
$6d \ ^2D_{5/2}$  & $-43.16$ &               &  25.75   &          &   & $-54.30$  &  & $-13.14$ & \\
\vspace{0.4mm}
$8s \ ^2S_{1/2}$  & 1578.99  &1577.8(23)$^e$ &          &          &   & 1986.13  &  &        &   \\
\vspace{0.4mm}
$8p \ ^2P_{1/2}$  & 294.97   &               &          &          &   & 371.08  & 373.0(1)$^b$ &        &   \\
\vspace{0.4mm}
$8p \ ^2P_{3/2}$  & 25.68    &               &  16.47   &          &   & 32.34   & 32.8(1)$^b$ &  $-8.40$  & $-7.7(9)^b$ \\
\vspace{0.4mm}
$7d \ ^2D_{3/2}$  & 22.67    &22.3(5)$^f$    &  5.98     &   Assume 0$^f$  &   & 28.52   &  &  $-3.05$  & \\
\vspace{0.4mm}
$7d \ ^2D_{5/2}$  & $-11.95$ & $-17.8(8)^f$   &  7.82    &64(17)$^f$&   & $-15.03$  &  &  $-3.99$  & \\
\vspace{0.4mm}
$9s \ ^2S_{1/2}$  & 623.50   &622.25(36)$^g$ &          &          &   & 784.26  &  &         & \\
\vspace{0.4mm}
$10s \ ^2S_{1/2}$ & 333.32   &               &          &          &   & 419.27  & 401(5)$^h$ &      & \\
\vspace{0.4mm}
$8d \ ^2D_{3/2}$  & 10.62    &               & 2.68     &          &   & 13.36   &13.0(6)$^h$  & $-1.37$  & Assume 0$^h$ \\
\vspace{0.4mm}
$8d \ ^2D_{5/2}$  &$-5.05$     &               & 3.49     &          &   & $-6.35$  & $-7.1(6)^h$  & $-1.78$  & $-2(10)^h$ \\
\vspace{0.4mm}
$9d \ ^2D_{3/2}$  & 5.77     &               & 1.42     &          &   & 7.26   &7.1(7)$^h$   & $-0.72$ & Assume 0$^h$ \\
\vspace{0.4mm}
$9d \ ^2D_{5/2}$  &$-2.59$     &               & 1.83     &          &   & $-3.25$  & $-3.6(4)^h$  & $-0.94$ & Assume 0$^h$ \\
\vspace{0.4mm}
$11s \ ^2S_{1/2}$ &168.85    &               &          &          &   & 212.39  & 225(3)$^h$   &      & \\
\end{tabular}
\end{ruledtabular}
\label{tab8}
$^a$\cite{coc}, $^b$\cite{duong}, $^c$\cite{coc1}, $^d$\cite{grossman}, $^e$\cite{simsarian}, 
$^f$\cite{grossman1}, $^g$\cite{gomez}, $^h$\cite{arnold}.
\end{table*}

As has been stated before, evaluation of the theoretical results for $A_{hf}$ and $B_{hf}$ require a knowledge of $g_I$
(i.e. $\mu_I$ and $I$) and $Q$ values of the atom. Our calculations using our CCSD$_{t3}$ method are the most rigorous 
theoretical results to date as they take into account more physical effects than previous calculations. 
The current best value of $\mu_I$
for $^{210}$Fr was extracted by combining the experimental $A_{hf}$ value of its $9S$ state with the 
corresponding calculation using the SDpT method \cite{gomez}. However, if 
experimental results for hyperfine structure constants of any state are known to high
precision, extraction of nuclear moments from these measurements can be justified. In reality, most of 
the measured $A_{hf}$ values in $^{210}$Fr are known quite precisely among which the ground state result is 
the most accurate (see Table \ref{tab8}). To infer $g_I$ value for $^{210}$Fr from the $A_{hf}$ results, we 
take the mean value from the data obtained combining the measurements with their corresponding calculations 
using the CCSD$_{t3}$ method for all the states except for the $D_{5/2}$ states. The reason for not 
considering results of these states is that the correlation effects in these cases are more 
than 100\% while for other states the principal contributions come from the DF values. In this approach, we obtain 
$g_I=0.733765942$ which corresponds to $\mu_I=4.40(5)$ of $^{210}$Fr. This value is in accordance with its earlier reported 
values as $\mu_I=4.40(9)$ \cite{ekstroem} and $\mu_I=4.38(5)$ \cite{gomez}. Unlike the case of $A_{hf}$, 
only two experimental values for $B_{hf}$ in $^{210}$Fr have been reported, among which the result for the $7D_{5/2}$ 
state might have been overestimated given that the wave functions of the $D_{5/2}$ states have an extremely small 
overlap in the nuclear region. Note that $B_{hf}$ of the $7P_{3/2}$ state is 51(4) MHz. Thus, combining the $B_{hf}$ value of the 
$7P_{3/2}$ state with the corresponding calculation, we obtain $Q=0.196(15)b$, where the uncertainty only from the 
measurement is taken into account, and the value estimated earlier was $Q=0.19(2)b$ \cite{ekstroem} for 
$^{210}$Fr. The agreement between these two values is because the same experimental $B_{hf}$ value has been used 
in both the results. By substituting these revised $\mu_I$ and $Q$ values, we have evaluated the 
theoretical $A_{hf}$ and $B_{hf}$ values of $^{210}$Fr and they are reported in Table \ref{tab8}. 

Keeping in mind the small differences among the calculations of the values of $A_{hf}/g_I$ and $B_{hf}/Q$ between 
$^{210}$Fr and $^{212}$Fr, as was mentioned earlier, we expect to observe the ratios between the $A_{hf}$
and $B_{hf}$ values from $^{210}$Fr and $^{212}$Fr for any given state to be almost equal to ratios of their
$\mu_I$ and $Q$ values, respectively, as per the demonstration in \cite{bijaya}. Considering all the 
experimental values known for the common states in both the isotopes, we find $g_I( ^{212}\text{Fr})/g_I( ^{210}\text{Fr})=1.25(1)$ 
and $Q( ^{212}\text{Fr})/Q( ^{210}\text{Fr}) = -0.51(5)$. Experimental results for
$A_{hf}$ and $B_{hf}$ are reported for more states in $^{212}$Fr than $^{210}$Fr. Excluding $A_{hf}$ 
results for the $D_{5/2}$ states owing to the reason stated previously, we obtain 
$g_I=0.923070701 $ for $^{212}$Fr when we combine its experimental $A_{hf}$ values of the remaining states with 
their respective $A_{hf}/g_I$ calculations. This corresponds to $\mu_I( ^{212}\text{Fr})=4.61(4)$. This, 
again, agrees with the previously reported value $\mu_I( ^{212}\text{Fr})=4.62(9)$ \cite{ekstroem}. In a 
similar procedure, we get $Q( ^{212}\text{Fr})=-0.10(1)b$ which is same as that given in \cite{ekstroem}. From 
these theoretical results, we get $g_I( ^{212}\text{Fr})/g_I( ^{210}\text{Fr})=1.26$ and 
$Q( ^{212}\text{Fr})/Q( ^{210}\text{Fr}) \approx -0.51$, which are in reasonable agreement with the above
estimated values from the measurements.  
 
 Comparison of the theoretical and experimental $A_{hf}$ and $B_{hf}$ results quoted in Table \ref{tab8} are quite 
satisfactory for almost all the states, but we find a large discrepancy between the experimental and
theoretical $B_{hf}$ values in the $7D_{5/2}$ state. This result requires further theoretical and experimental 
verifications. Also, it was assumed that $B_{hf}$ values of other $D_{3/2,5/2}$ states were negligible while
extracting the experimental $A_{hf}$ values of the corresponding states, however the present work shows that $B_{hf}$ of the 
$7D_{3/2}$ state is about 6 MHz. 
 
\section{Conclusion}

In summary, we have employed many-body methods at different levels of approximation to study the magnetic
dipole and electric quadrupole hyperfine structure constants of the first 17 states in $^{210}$Fr and 
$^{212}$Fr. This work demonstrates the importance of the inclusion of the non-linear terms in the 
coupled-cluster method, that accounts for the contributions from the triples and quadrupole excitations, for 
the accurate evaluation of the above quantities and for the studies of the violations of parity and time reversal 
symmetries in Fr. By combining the experimental values with our corresponding 
calculations, we obtain $\mu_I=4.40(5)$ and $Q=0.196(15)b$ for $^{210}$Fr and $\mu_I=4.64(4)$ and 
$Q=0.10(1)b$  $^{212}$Fr. A reasonably good agreements between the theoretical and experimental results 
are obtained except for the electric quadrupole hyperfine structure constant of the $7D_{5/2}$ state. 
Theoretically predicted values for the hyperfine structure constants of many states including the $6D$ 
states, in the present work, could be tested experimentally in the future.

\section*{Acknowledgement}
This work was supported partly by INSA-JSPS under project no. IA/INSA-JSPS Project/2013-2016/February 28,2013/4098. 
Computations were carried out using the 3TFLOP HPC cluster at Physical Research Laboratory, Ahmedabad.

\end{document}